# Is the edge really necessary for drone computing offloading? An experimental assessment in carrier-grade 5G operator networks

David Candal-Ventureira | Francisco Javier González-Castaño | Felipe Gil-Castiñeira | Pablo Fondo-Ferreiro

atlanTTic, University of Vigo, Information Technologies Group, Vigo, Spain

**Correspondence**
Francisco Javier González-Castaño, EI Telecomunicación, Campus Universitario Lagoas-Marcosende, 36310 Vigo, Spain.
Email: javier@det.uvigo.es

**Funding information**
Ministerio de Ciencia e Innovación, Grant/Award Numbers: PDC2021-121335-C21, PID2020-116329GB-C21, PRE2021-098290

**Abstract**

In this article, we evaluate the first experience of computation offloading from drones to real fifth-generation (5G) operator systems, including commercial and private carrier-grade 5G networks. A follow-me drone service was implemented as a representative testbed of remote video analytics. In this application, an image of a person from a drone camera is processed at the edge, and image tracking displacements are translated into positioning commands that are sent back to the drone, so that the drone keeps the camera focused on the person at all times. The application is characterised to identify the processing and communication contributions to service delay. Then, we evaluate the latency of the application in a real non standalone 5G operator network, a standalone carrier-grade 5G private network, and, to compare these results with previous research, a Wi-Fi wireless local area network. We considered both multi-access edge computing (MEC) and cloud offloading scenarios. Onboard computing was also evaluated to assess the trade-offs with task offloading. The results determine the network configurations that are feasible for the follow-me application use case depending on the mobility of the end user, and to what extent MEC is advantageous over a state-of-the-art cloud service.

## 1 | INTRODUCTION

Fifth-generation (5G) networks have been designed to support many new use cases with their own very specific performance requirements. Consequently, multiple new technologies have been introduced both into the core network and the radio interface of these cellular networks. Of special relevance are the softwarisation and virtualisation of their components. Nowadays, elements that were traditionally implemented on proprietary hardware black boxes run as software on

---

**Abbreviations:** MEC, multi-access edge computing; QoS, quality of service; SDN, software-defined networking; UAV, unmanned aerial vehicle.

top of general purpose computing servers. Operators can efficiently manage the life cycle of the virtualised cellular network entities, by deploying or withdrawing instances and allocating resources to them depending on dynamic end user demands. The software-defined networking (SDN) paradigm has enabled new network architectures that are more flexible and easier to manage. This paradigm also provides operators with broad and complete information about the status of their networks and supports globally consistent policies.

Virtualisation and SDN have also allowed the efficient integration of the new remote multi-access edge computing (MEC) paradigm for mobile networks.[1] Operators can lease the MEC resources at the edge of their networks as an alternative to cloud computing services, and fulfil the service level agreements (SLAs) with mobile users. Edge computing is advantageous because it brings application instances close to end users: since the traffic is constrained to the edge, communication latency drops, bandwidth usage improves owing to spatial reuse, and privacy and integrity are easier to guarantee. MEC has posed diverse research challenges, such as task partitioning between onboard resources and MEC servers,[2] depending on local and remote computing capacities and quality of service (QoS) demands. In this work we consider a use case in which an unmanned aerial vehicle (UAV) application is partitioned into computationally intensive image processing tasks, which are offloaded to the edge, and minor supporting tasks, such as image capturing and the drone commander, that are run onboard. This type of vehicular applications must satisfy the stringent requirements of autopiloting and video analytics services.[3-5] Edge computing is specially interesting for UAV task offloading, as UAVs generally have scarce computational capabilities due to weight, size, and battery limitations.[6]

In this article we evaluate the first experience of drone computing offloading in real commercial and private carrier-grade 5G networks. A drone follow-me service has been implemented as a representative use case of remote video analytics. In this application, an image of a person captured by a drone camera is processed at the edge, and image tracking displacements are translated into positioning commands that are returned to the drone for it to continuously keep the camera focused on the person. We describe and characterize the application to identify the different processes that contribute to overall service delay. We then study the maximum acceptable service delay depending on user speed, and we analyse the service delay in different scenarios including computing offloading to 5G commercial non standalone (NSA) and private carrier-grade standalone (SA) networks. In addition, we evaluate onboard computing to assess the trade-offs with computing offloading. The analysis was performed on an edge platform of a real operator, an real implementation of an edge platform of an ideal private network, a widely used commercial cloud platform and, as a reference for comparison with previous research, a Wi-Fi wireless local area network (WLAN). Our results suggest the network configurations that are feasible for different usage scenarios of the remote follow-me service depending on the mobility of the end user, and the advantages of MEC compared to a state-of-the-art cloud service.

The rest of this article is structured as follows: Section 2 discusses related work. Section 3 introduces the follow-me service. Section 4 describes the architecture of the solution and the setup for the experimental evaluation. Section 5 evaluates service feasibility in real 5G networks and a WLAN, both with MEC and cloud support, and in case of onboard computing. Finally, Section 6 concludes the article.

## 2 | RELATED WORK

The MEC paradigm has been extensively studied in the literature. One of the its challenges is the scheduling of computing resources either at the network edge or the cloud to serve incoming requests. This scheduling must take into account the availability and proximity of edge computing resources, the requirements of the services (which may change even during the same user session), the mobility and spatial distribution of the users and so forth. As a result, substantial work on the MEC paradigm has tackled this challenge, for example by optimising offloads to minimise the response time of the services;[7-10] or by seeking trade-offs between quality of experience (QoE) and service deployment costs[11] and the edge computing infrastructure these services require.[12] Other lines of research are the design of architectures for dynamic deployment of services on an edge computing scenario;[13-15] and task distribution among local and remote heterogeneous servers.[16] There also exists abundant literature on the performance of computing services depending on the relative location of the servers with respect to the users (e.g., network edge vs. cloud).

Regarding theoretical work, diverse aspects have been thoroughly covered.[17] Maheshwari et al. developed an analytic model of the response times of a service task running on cloud or edge servers.[18] It considers processing time requirements, the load of the server and the bandwidth of the network links between the server and the end user. The results suggest that response time decreases by adding up computational resources and network bandwidth, but the improvement is marginal beyond certain link capacity and may even worsen if the computational capacity of the servers is too high compared to the bandwidth of the links. A simulator of edge computing applications modelling both the computational and networking systems was presented in Reference 19.

In the specific case of vehicular scenarios, different works[3,4,20-22] have studied the performance of drone-assisted and road safety applications. For example, Hayat et al. studied an offloaded multi-state constraint Kalman filter for visual-inertial drone navigation (MSCKF-VIO).[3] They modelled the response time of the service for different image resolutions in three scenarios: onboard processing, full, and partial offloading. The results show the importance of uplink channel rate: even if the remote server has much more computational resources than the drone, the transmission time of the images can be an issue. Therefore, onboard pre-processing to extract the image features to be sent to the offloading service instead of the original image may be beneficial to minimise transmissions. Messous et al. envisage a scenario where several drones execute a heavy task or offload it to the edge servers of a cellular network or a Wi-Fi basic service set (BSS).[20] When too many drones decide to offload their tasks to the same server, service time may grow significantly due to communications or server saturation. Therefore, the authors propose a game theory algorithm for determining the locations where the tasks should be executed. They demonstrate by simulation that their approach is better on average than centralised computing. In Reference 21, the same authors evaluate the response time of a service that is split into subtasks and distributed among different nodes according to their computational and networking capabilities. The results show that optimal node selection reduces response times significantly. Bylykbashi et al. describe two remote monitoring models for alerting drivers based on car sensors.[4] They study the latency of this service with edge and fog computing, but no theoretical nor experimental evaluations based on network conditions are presented. Other works have compared edge and cloud deployments of applications for collision detection and vulnerable road user (VRU) warning.[22-24]

Regarding experimental work, different authors have analysed the performance of latency-aware applications.[6,25-27] Motlagh et al. evaluate the processing time and energy consumption of a facial recognition service based on a video stream captured by a drone, by comparing onboard computing with computational offloading to the network edge.[6] They demonstrate that offloading can save drone battery while improving service performance. This is relevant for this work because it backs the interest of drone service offloading. Takagi et al. evaluate the impact of latency in QoE.[26] Their study is based on a mixed-reality remote application to control a robot with the gestures of an user, who receives video feedback from the robot. The video is processed by a remote server that adds overlay information to the video before streaming it to the user. All nodes are interconnected by a Wi-Fi BSS, and synthetic delay is injected at the network interface of the server (that is, even though the study is experimental, edge computing is emulated). Dautov et al. evaluate the performance of a face detection and recognition application that is offloaded either to nodes nearby or to the cloud.[27] There also exists work on object detection and tracking applications based on computer vision[28-30] that are offloaded from a drone to a remote server. However, all communications in these works take place through Wi-Fi WLANs. To the best of our knowledge no previous experimental research has considered similar offloading scenarios in a real 5G cellular network with a real 5G MEC platform. Consequently, some relevant aspects beyond radio access technology have not been considered by previous works, such as the impact of the core network modules, the virtualised architecture that provides remote computing services, and the softwarised network. In fact, most previous authors consider that "edge computing" is any architecture that takes computing close to the radio access segment regardless of its radio network technology (some examples include LoraWan[31] and LTE[32]). This may be conceptually valid and interesting, but the results of these experiences cannot be extrapolated to 5G MEC. We remark that, even though our previous study[33] measured the delay and data rate of actual 5G data transmissions, it was limited to communications between an end device and the 5G and LTE cores of a laboratory testbed.

Summing up, previous works have typically emulated edge computing with a server in the same WLAN as the user terminal, or emulate the backhaul in a laboratory environment[34] at most. In fact, the lack of real large-scale environments has led to the development of specialised simulators.[19] In this work, we fill this gap by evaluating real MEC systems that are integral parts of real-world 5G operator networks. This allows us assessing the feasibility of edge offloading for the use case under evaluation.

Thus, the main contribution of this article is an experimental assessment of a representative use case of drone computing offloading to real edge systems that are associated to carrier-grade 5G Next Generation NodeBs (gNodeBs) and core networks, which are accessed through commercial off-the-shelf (COTS) end devices with real 5G subscriber identity module (SIM) cards. The goal is to determine if a the use case is feasible by jointly analysing the effect of the 5G radio access network (RAN), the backhaul, the core network and the edge virtualisation architecture. Besides, real 5G MEC platforms are compared with a commercial cloud for the same purpose.

To the best of our knowledge no previous work has experimentally evaluated a drone follow-me application at the edge, so the practical evaluation of this particular computer vision application is a side contribution of our work.

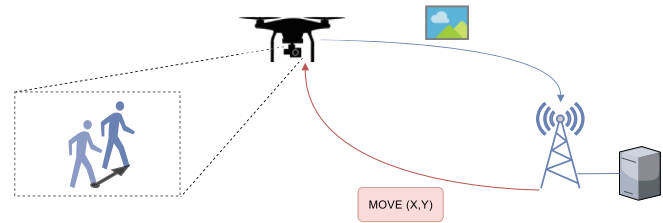

**FIGURE 1** High level representation of the follow-me service.

## 3 | DESCRIPTION OF THE APPLICATION

We consider a follow-me application that guides the drone by tracking the movements of a person. This is considered a representative use case of drone information processing that could be offloaded as a service to a MEC platform of a real 5G operator network. The motivation of this scenario is to take image processing, which is a computationally expensive task, outside the drone limited resources, while keeping the service delay below some bound. Other benefits of an edge deployment are the capability to update the follow-me application with new features without reprogramming the firmware of the drone, and the possibility to use the video stream for training the artificial intelligence (AI) models in the background.

Figure 1 illustrates the proposed use case. The follow-me application detects the user in the frames captured by the camera of the drone. Based on this information, the application determines the command to be sent to the drone to track the user.

In each captured frame, the follow-me application detects the position of the user and, based on his/her pre-configured height, the fixed height of the drone and the focal length and aperture of the camera, the distance to the user is estimated. Then, the application determines the movement that the drone must perform, which consists of two components, an x-axis movement (forward or backward) to keep the distance to the user, and a y-axis movement (left or right), so that the user is kept at the centre of the frame. This is translated to the two-dimensional coordinate where the drone must move from its current position upon reception of the corresponding command.

As previously noted, drones usually have limited computational resources due to payload and battery constraints. For this reason, it may be desirable to offload computationally intensive applications to external servers through some wireless communication technology (such as Wi-Fi, LTE, and 5G New Radio). This may greatly improve the performance of drone applications compared to onboard execution, as demonstrated by previous works.[6] Nevertheless, service delay (including communication delay) should not exceed the limits imposed by the requirements of the application.

In this scenario, by assuming that computational offloading is desirable, we are interested in comparing different implementations on a real 5G edge and a commercial cloud platform. Moreover, we wish to compare a commercial shared edge setup, in which processing nodes are connected to an operator backhaul network at best and serve several gNodeBs, with a dedicated private edge setup, where data plane and edge computing resources are close to the end user location or even at the serving gNodeB itself.

## 4 | ARCHITECTURE OF THE SOLUTION

### 4.1 | Software architecture

Figure 2 illustrates the software architecture of the follow-me application in the testbed. Green boxes have been specifically implemented for this research, whereas the red box represents the proprietary remote control application programming interface (API) of the drone. Gray boxes represent low-level proprietary functions of the drone, which are controlled through the remote control API. The service is partitioned into two main software components:

- The client application at the drone side. This lightweight application is in charge of capturing images and streaming them to the server application and of executing drone commands. On system start-up, it initiates a drone take-off procedure. Then, the client application starts streaming the video to the server application and listens to incoming commands from the latter. The client application interacts with the remote control drone API according to the received positioning commands.

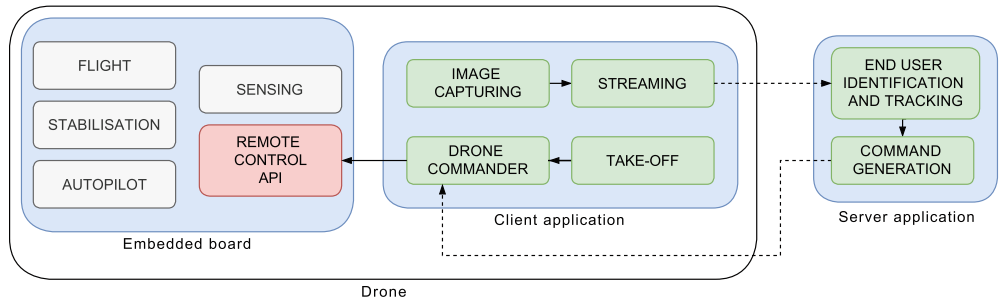

**FIGURE 2** Software architecture of the proposed solution.

- The server application at an external server (in our case, a carrier-grade MEC platform or a commercial cloud). It handles all computationally complex tasks that are offloaded from the drone. It receives the video streaming from the client application and identifies and tracks the end user in the frames as explained in Section 3.

## 4.2 | Communications architecture

Figure 3 illustrates the different network architectures in the testbed. The drone and the remote server (MEC or cloud) are interconnected through a 5G network. The testbed considers two different 5G network setups:

- Commercial NSA 5G network setup. A commercial NSA 5G network deployed by a multinational mobile operator in Vigo, Spain. The performance is not deterministic, since the network resources are shared with external users outside the experiment.
- Private carrier-grade SA 5G network setup. A private SA 5G network deployed within University of Vigo premises using carrier-grade equipment donated by a multinational mobile operator, including commercial frequencies that were temporarily granted on an exclusive basis. This network is isolated from commercial users, both at the radio interface and the core network. Since there is no interference from other users, this setup performs as an ideal SA 5G network.

Both setups operate in the n78 band (3.3–3.8 GHz). We evaluate in each setup the deployment of the target application at the MEC and the cloud. In the MEC case, the offloading server is closer to the 5G data plane core network functions, whereas in the cloud case a state-of-the-art commercial service is used. Note that, as depicted in Figure 3, MEC servers are semi-centralised in the commercial network setup. That is, they are placed in a few zones of the coverage area of the operator, so that each server is shared by all the base stations within the corresponding zone. Therefore, a base station may be substantially distant from its MEC server from a networking point of view, resulting in increased communication latency compared to an ideal MEC scenario. Conversely, in the private network setup, the MEC server is co-located with the base station.

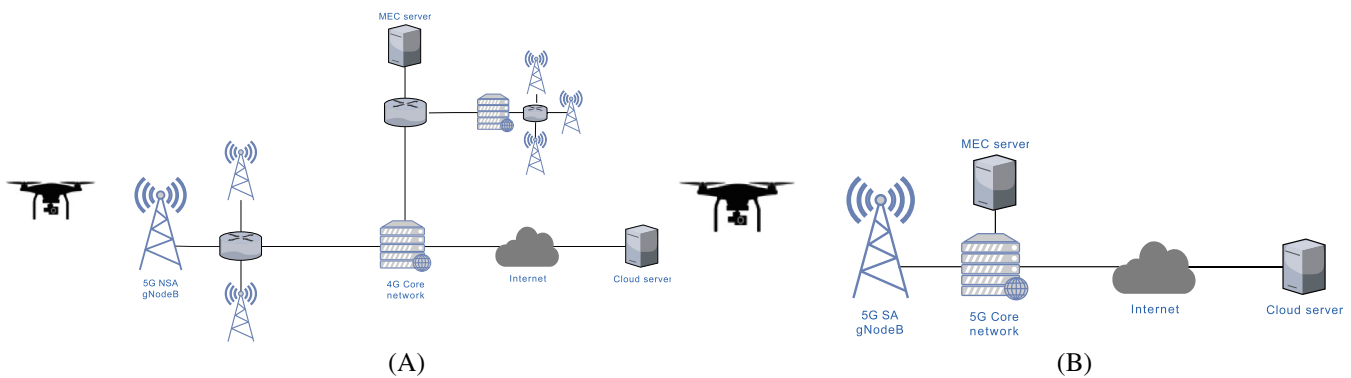

**FIGURE 3** 5G network architectures of the testbed. (A) Commercial NSA 5G network setup; (B) private carrier-grade SA 5G network setup

## 4.3 | Implementations

Figure 4 represents the hardware architecture of the user end side. It is composed by a commercial DJI Matrice 210 v2[*] drone, which is controlled by an onboard single-board computer (SBC) that is wired to the drone electronics. The SBC is a Raspberry Pi 4 Model B[†] with a quad core Cortex-A72 (ARM v8) 64-bit system on a chip (SoC) @1.5GHz with 2 GB RAM. The SBC is connected to the drone electronics through a transistor–transistor logic (TTL) port for command transmission via universal asynchronous receiver/transmitter (UART) serial communications and to the external power port of the drone for power supply. The SBC is equipped with a 5G user equipment (UE) dongle and a web camera. The 5G UE dongle is a 3GPP Release 15 compliant Quectel 5G RM500Q-AE[‡], which supports MIMO $4 \times 4$ and is compatible with a wide range of SA and NSA Sub-6 GHz 5G bands. The camera is a Logitech StreamCam[§] web camera with vertical and horizontal angles of vision (AOVs) VAOV = 41.2° and HAOV = 67.5°, capturing $1280 \times 720$ images at 30 frames per second.

Both the client and server applications were implemented in C++. The client application generates the video stream with the GStreamer[¶] multimedia framework, and employs the C++ DJI Onboard Software Development Kit (OSDK) V4.0.0 API to control the drone. The server application was deployed as a Docker container. It respectively uses the OpenVINO person-detection-retail-0013[#] and person-reidentification-retail-0031[‖] pre-trained deep learning models for detecting people in image frames and tracking them with regard to previous frames. Positioning commands are sent to the drone through a TCP session.

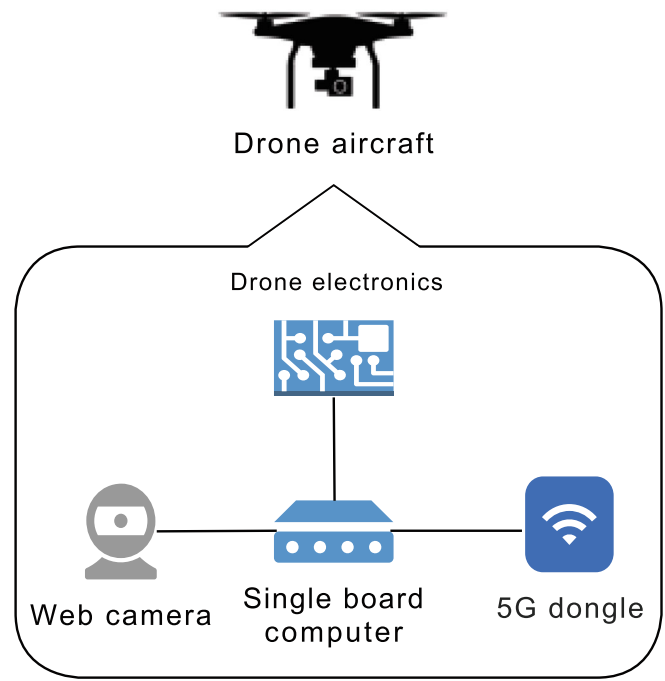

**FIGURE 4** User end side hardware architecture.

[*]https://www.dji.com/es/matrice-200-series-v2
[†]https://www.raspberrypi.com/products/raspberry-pi-4-model-b/
[‡]https://www.quectel.com/product/5g-rm500q-ae
[§]https://support.logi.com/hc/es/articles/360042528854-StreamCam-Technical-Specifications
[¶]https://gstreamer.freedesktop.org/
[#]https://docs.openvino.ai/2021.2/omz_models_intel_person_detection_retail_0013_description_person_detection_retail_0013.html
[‖]https://docs.openvino.ai/2019_R1/_person_reidentification_retail_0031_description_person_reidentification_retail_0031.html

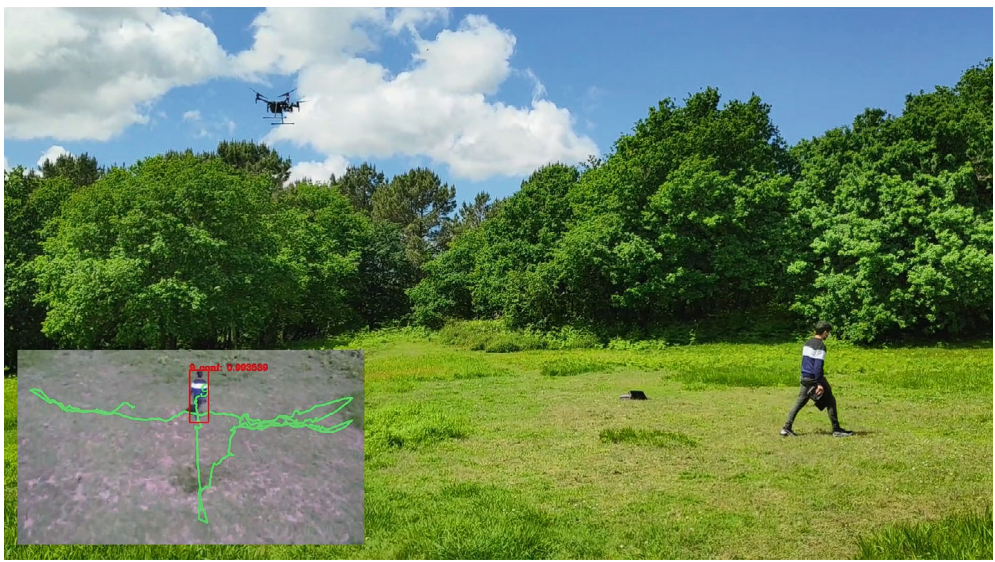

**FIGURE 5** Illustration of follow-me system in use.

Figure 5 illustrates the testbed application in use, with an overlay image (bottom left) showing the most recent frame captured by the drone as processed by the remote platform. In this frame, detection data was used to draw a rectangle around the final user position, and previous detection data to draw the tracking path of the user.

Table 1 shows the main specifications of the 5G network setups in this work. As previously said, the commercial network was NSA at the time this article was written, that is, the 5G gNodeB was connected to a 4G core network. Conversely, the carrier-grade private network is a SA setup, meaning that both the gNodeB and the core network employ full 5G technologies. The bandwidth of the commercial network is slightly higher (60 MHz vs. 50 MHz) and supports a more complex MIMO scheme, although a single client (as in our case) can only exploit this scheme if its modem can handle the corresponding number of beams. However, the 5G dongle we used in both network setups only supports up to $4 \times 4$ MIMO, so the setups are comparable in this regard. Both network setups operate in the n78 band and have the same subcarrier spacing and TDD frame structure configuration. The main difference between the two gNodeBs is that, while the commercial gNodeB creates a typical macrocell from a mast to cover a large area, the private gNodeB is a microcell that only covers its close surroundings.

An Intel Smart Edge Open** platform was deployed at the university facilities on a server that was immediately close (from a networking point of view) to the private 5G core network. In the commercial setup, the operator gave us access to its proprietary edge platform to evaluate the target application. Note that, as shown in Figure 3A, in this case the MEC server is semi-centralised, in the sense that it is shared by diverse gNodeBs. That is, it is significantly farther away from any given gNodeB than in the private network setup, in which the core, the RAN and the MEC platform belong to the same physical location. The delay to the commercial MEC platform should be necessarily higher as a result. To evaluate the cloud scenario we chose to deploy the remote service on an Azure node. Region "France Central" (based on Paris) was selected for the deployment, since it was the data centre that was geographically closest to our premises yielding the best raw network performance in terms of latency and throughput. In all these three offloading platforms (private MEC platform, commercial MEC platform and commercial cloud) the follow-me service was deployed as a container through Kubernetes. Table 2 shows the specifications of the servers in the platforms. Their differences have been taken into account in the evaluation by decoupling transmission latency from processing delays.

**TABLE 1** Parameters of the 5G network setups.

|  | Network deployment | Band | Bandwidth | MIMO | Subcarrier spacing | TDD frame structure | Coverage type |
| --- | --- | --- | --- | --- | --- | --- | --- |
| Commercial | NSA | n78 | 60 MHz | $64 \times 64$ | 30 kHz | DDDSU, 2.5 ms S: 10D, 2U, 2GP | Outdoor |
| Private | SA | n78 | 50 MHz | $4 \times 4$ | 30 kHz | DDDSU, 2.5 ms S: 10D, 2U, 2GP | Indoor |

**https://www.intel.com/content/www/us/en/developer/tools/smart-edge-open/overview.html

**TABLE 2** Specifications of the MEC and cloud computing platforms.

|  | Architecture | CPU frequency | Threads | RAM |
| --- | --- | --- | --- | --- |
| Intel Smart Edge Open | x86-64 | 2.50 GHz | 8 | 4 GB |
| Edge platform of mobile operator | x86-64 | 2.30 GHz | 8 | 4 GB |
| Azure | x86-64 | 2.60 GHz | 4 | 4 GB |

## 5 | EXPERIMENTAL EVALUATION

This section evaluates the use case in the different 5G network setups with their respective MEC and cloud computing platforms, as described in Section 5. We first present the evaluation methodology in Section 6.1. Then, Section 6.2 evaluates the performance of the different 5G network setups in terms of latency and throughput. Next, the follow-me service is evaluated in Sections 6.3–6.5 in terms of end-to-end service delay. Finally, Section 6.6 discusses the feasibility of the results by considering the target requirements.

### 5.1 | Methodology

The testbed in Section 4 was deployed in a controlled environment where the carrier-grade gNodeB of the private 5G network setup is installed. The testbed application was configured to process images of a person moving around a given area. The drone stayed at a fixed height of $h = 6$ meters above the ground. The initial position of the person was $d = 10$ m away from the vertical of the drone.

Assuming first that the drone is static and points to the user, and that the user moves (*i*) towards one side of the camera frame (*ii*) towards the camera or (*iii*) away from the drone (that is, direction *i* is perpendicular to directions *ii* and *iii* and directions *ii* and *iii* are opposite):

(i) By applying simple trigonometric calculations, the real distance between a centred user and a lateral side of the camera frame is $W = \sqrt{d^2 + h^2} \cdot \tan(\text{HAOV}/2) = 7.79$ m.
(ii) The maximum distance the user can walk towards the drone within the frame is $d - L_2 = d - h \cdot \tan(\arctan(d/h) - \text{VAOV}/2) = 5.24$ m.
(iii) The maximum distance the user can walk away from the drone within the frame is $L_1 - d = (h - h_u) \cdot \tan(\arctan(d/h) + \text{VAOV}/2) - d = 13.51$ m.

Where $h_u$ is the height of the user and $L_1$ and $L_2$ are the distances from the camera to the limits of the field where the user's body is fully captured by the camera, as shown in Figure 6B. We considered a user height $h_u = 1.7$ m in our calculations. Note that we are supposing that the whole body of the user must be inside the image to recognise and track him/her. This is, however, conservative, since the system can work when the user is partially inside the image.

Therefore, the shortest distance for the user to leave the frame is when moving towards the camera, 5.24 m. We set the design goal of making at least one correction before the user moves one third of that distance, 1.75 m. We will consider three scenarios: an average person walking at 5 km/h or riding a bicycle at 25 km/h (as a representative case of fast passive assistance), with a typical dash running speed of 10 km/h lying in the middle. In the worst case, assuming that the person at the centre of the frame is already moving towards the motionless drone, the person would cover one third of the distance to the edge of the frame in 1.26 s if walking, 0.63 s if running and 0.25 s if riding a bicycle. For the motion estimation to be feasible in each of these situations, at least three frames should be captured before the corresponding deadline, so that the velocity and acceleration of the user can be calculated. Therefore, the maximum total service delay to guarantee the feasibility of the use case would be $t_w = 419.2$ ms for a walking person, $t_r = 209.6$ ms for a runner and $t_b = 83.84$ ms for a cyclist.

The service delay for the target application (that is, the time it takes since the onboard SBC sends an image until the corresponding positioning command is received from the server application) is composed of three delays:

- Image transmission delay: The time to transmit the image to the server. Each image is sent as a burst of UDP packets.

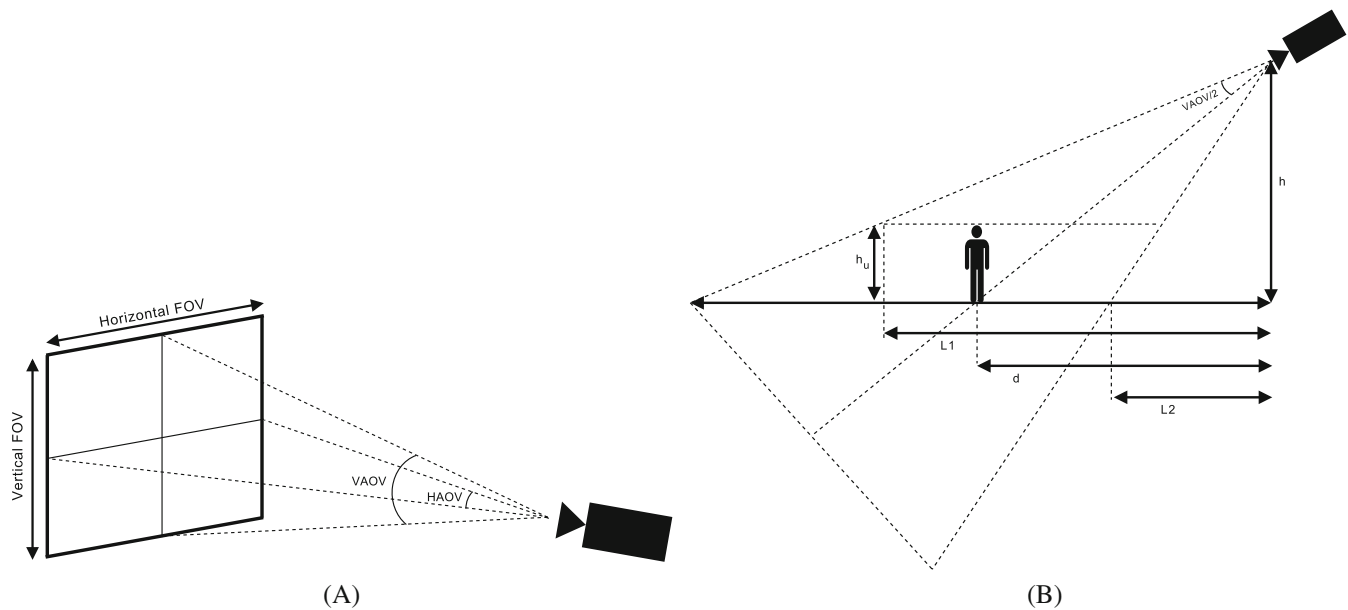

**FIGURE 6** Geometry of the vision fields of the camera. (A) Horizontal and vertical AOVs and Fields of Vision (FOVs); (B) vertical camera plane.

- Processing delay: The time the server takes to process the image to identify and track the end user and compute the corresponding positioning command.
- Command transmission delay: The time to transmit the command from the server to the onboard SBC. Commands are transmitted in a TCP connection. The client device acknowledges each command transmission upon reception.

The video stream carries 720p images at 30 frames per second, resulting in a maximum uplink rate of 2.095 Mbps. The data payload of each command transmitted from the server to the drone is 8 bytes, corresponding to a data rate of 17.344 kbps in the downlink channel and 15.468 kbps in the uplink channel for command acknowledgments, both at the maximum frame rate. Note that, since command packets are lightweight transmissions with a data payload of only 8 bytes, the transmission times of a command and its acknowledgment are very similar, especially in high capacity networks such as those in the testbed.

The processing delay was measured in the server as the time interval since the reception of a frame to the moment when the corresponding command was ready to be transmitted back. Then, the overall service delay was computed at the client as the time since the beginning of the transmission of a frame until the respective command was received. This was accomplished by comparing the timestamps of the packets that were exchanged by the client and the server, from a network trace captured at both sides with the *tcpdump* packet analyser tool. Since the shortest possible delays can be in the order of a few milliseconds, they must be measured by comparing the timestamps of events that happen within the same host, because clock synchronization protocols such as Precision Time Protocol (PTP) cannot guarantee such a precision and solutions such as GPS-based clock calibration were not available. Consequently, the command transmission delay was estimated at the server side by halving the time since sending a command until receiving the client acknowledgment. This measurement was conducted, as at the client side, by analysing a network trace captured at the server. Finally, the image transmission delay was computed by subtracting the command transmission delay and the processing delay from the overall service delay.

In our evaluation, for each network and offloading platform setup, end-to-end overall service delay was evaluated from 12,000 samples of each measurement described in the preceding paragraph.

## 5.2 | Performance analysis of 5G network setups

The latency and throughput of the 5G networks between the drone SBC and the MEC and cloud platforms were measured with the *ping* and *iperf3* tools widely used in network performance evaluation tests.[35,36] The *ping* command was configured with the default payload of 56 B and interval of 1 s between transmissions. The *iperf3* tool was configured to use four parallel UDP streams with an aggregated bitrate of 450 Mbps for the downlink channel and 150 Mbps for the uplink channel. These two bitrates were set to exceed the estimated maximum data rate of the networks to compute their capacity. We took 1000 measurements of the latency and the throughput for each combination of a particular 5G network setup with a particular remote computing platform. For the throughput evaluation, the first 50 values reported by *iperf3* were discarded to select measurements in a stationary state. In the latency evaluation no measurements were discarded, but we verified that the PDU session was not interrupted before the evaluations, so that there was no extra latency in the first results due to PDU session reestablishment. In the following tables and figures, we refer to the commercial NSA 5G network setup as "commercial network" and to the private carrier-grade SA 5G network setup as "private network".

Table 3 shows the round-trip time (RTT) delay to the MEC and cloud platforms in the 5G network setups under evaluation. The private network, as expected, was much more predictable in terms of network latency than the commercial network. Note that commercial network results are affected by the coexistence with external users (outside the experiment) using the same network, whereas the private network was fully dedicated to the drone UE. Moreover, the results show that the distance between the base station that is serving the user and the computing server had substantial impact in latency in our use case. As described in Section 5, the MEC platform of the private network is deployed in the same facilities as its base station and its core. On the contrary, the MEC platform of the commercial network is considerably farther away from the base station. Even though data traffic can be directly forwarded by a single User Plane Function (UPF) node between these two elements, many forwarding elements may be involved in the commercial MEC scenario. Therefore, the latency between the end user and the MEC platform was 3.53 times higher than the private network scenario. Regarding the cloud evaluations, the difference between network setups is much smaller but still noteworthy. The average RTT delay of the private 5G SA network to the cloud was 63.8% of that of the commercial 5G NSA network. By assuming that the delay from the backbone of the two network setups to the cloud server is comparable, the radio fronthaul and the distance from the base station to the core network in the commercial network jointly account for an extra delay of 21 ms compared to the private network.

The 5G standard[37] determines that the maximum theoretical supported data rate for the downlink and uplink channels in 5G New Radio (NR) is given by Equation (1):

$$data\_rate = 10^{-6} \cdot \sum_{j=1}^{J} \left( v_{\text{Layers}}^{(j)} \cdot Q_m^{(j)} \cdot f^{(j)} \cdot R_{\max} \cdot \frac{N_{PRB}^{BW(j),\mu} \cdot SC_{RB} \cdot \overline{S}^{(j)}}{T_s^{\mu}} \cdot \left(1 - OH^{(j)}\right) \right) \text{ Mbps}. \quad (1)$$

where:

- $J$ is the number of aggregated 5G carrier components.
- $v_{\text{Layers}}^{(j)}$ is the maximum number of MIMO layers supported between the device and the base station for carrier $j$.
- $Q_m^{(j)}$ is the modulation order for carrier $j$.
- $f^{(j)}$ is the scaling factor for carrier $j$.
- $Rmax$ is the target code rate, or the ratio between non-redundant and total transmitted bits, which depends on the MCS of the UE.
- $N_{PRB}^{BW(j),\mu}$ is the maximum resource block (RB) allocation in bandwidth $BW$ with numerology $\mu$, for carrier $j$.
- $SC_{RB} = 12$ is the number of subcarriers per RB.

**TABLE 3** Network RTT delay measurements (ms).

|  | Average | Median | Minimum | Maximum | 10th percentile | 90th percentile | Std. deviation |
| --- | --- | --- | --- | --- | --- | --- | --- |
| Commercial to MEC | 40.560 | 37.8 | 35.1 | 61.4 | 36.0 | 49.34 | 5.433 |
| Commercial to cloud | 58.013 | 56.8 | 50.2 | 83.2 | 51.3 | 67.8 | 6.563 |
| Private to MEC | 11.504 | 10.8 | 9.19 | 18.8 | 9.71 | 14.8 | 1.981 |
| Private to cloud | 37.023 | 36.3 | 34.2 | 73.9 | 34.8 | 40.7 | 2.611 |

- $\overline{S}^{(j)}$ is the average number of OFDM symbols that are reserved for downlink (or uplink) data transmission, for carrier $j$.
- $T_s^\mu = \frac{1ms}{14 \cdot 2^\mu}$ is the average OFDM symbol duration in a subframe with numerology $\mu$, assuming normal cyclic prefix.
- $OH^{(j)}$ is the overhead for carrier $j$. This parameter depends on the frequency band range and the communication direction (downlink or uplink). In the FR1 region (submillimetre frequencies), $OH$ is 0.14 for the downlink and 0.08 for the uplink.

The networks in both setups have a single carrier component. Even though the 5G modem that was used in the experiments supports $4 \times 4$ MIMO for both the uplink and downlink channels according to its specifications, we found that the UE was actually using only 3 data layers for the PDSCH channel and only 1 data layer for the PUSCH in the private network experiments, and 2 data layers for each PDSCH and PUSCH channel when it was connected to the commercial network. This was corroborated by the answers of the 5G modem to AT command *AT+QNWCFG="nr5g_csi"*. Both network setups have a subcarrier spacing of 30 kHz, which corresponds to a numerology of $\mu = 1$. Therefore, the average OFDM symbol duration is $T_s^\mu = 35.71 \times 10^{-6}$ s. According to Reference 38, for $\mu = 1$, $N_{PRB}^{BW,\mu}$ is 133 PRBs for 50 MHz and 162 PRBs for 60 MHz. In the TDD frame structure in Table 1, three out of five subframes are reserved for downlink transmission, one is dedicated to the uplink channel and one is shared. Each shared subframe allocates 10 symbols to the downlink and 2 to the uplink. This results in an average number of OFDM symbols $\overline{S}$ of 0.7429 for the downlink and 0.2286 for the uplink, for both networks.

The UEs adapt their MCS dynamically to the conditions of the radio interface. Downlink and uplink channels are independently considered. AT command *AT+QNWCFG="nr5g_csi"* reports the MCS index for the downlink channel, and no AT command is available to read the MCS for the uplink channel. However, in our setup we observed that the downlink data rate of the modem dropped when a serial communication was opened, which invalidates the use of AT commands for our purposes. Therefore, we relied on the automatic report of the base station of our private network, whose metrics include precise MCS indices of the scheduled PRBs. We collected these during the throughput evaluation of the private 5G SA network. Figure 7 shows their distribution, which allows estimating the maximum theoretical data rate. Table 5.1.3.1-2 in 3GPP technical specification TS 38.214 version 16.2.0[39] shows the relation between the MCS indices plus the modulation order $Q_m$, on the one hand, with the target code rate $Rmax$, on the other. As shown in Figure 7, virtually all the PRBs that are scheduled for uplink transmissions used MCS indices 26 and 27, which correspond to a 256-QAM modulation. In the downlink the distribution of the MCS indices was more scattered, with 97% in [17,21]. By considering a scaling factor of 1, the theoretical maximum data rates of the private 5G SA network for the downlink and uplink channels were 428.229 and 67.842 Mbps, respectively. As we could not obtain the indices of the commercial 5G NSA network, we had to guess the maximum throughput from network measurements.

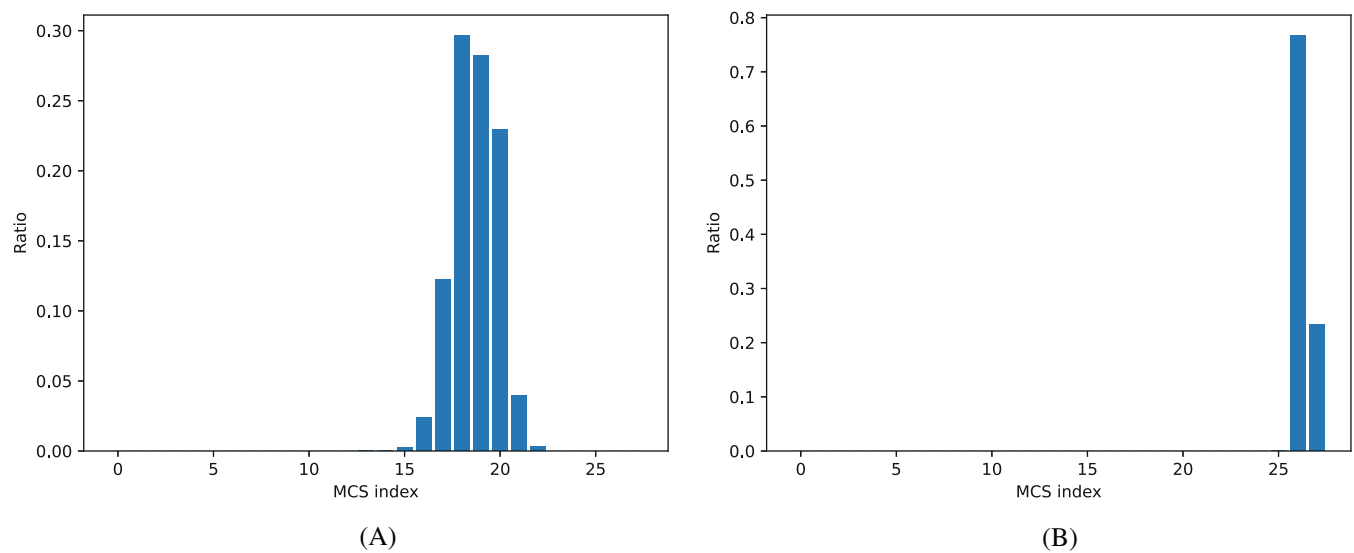

**FIGURE 7** MCS indices of private 5G SA network during throughput evaluation. (A) MCS indices for the downlink channel; (B) MCS indices for the uplink channel

**590** | WILEY | CANDAL-VENTUREIRA ET AL.**TABLE 4** Network downlink throughput measurements (Mbps).

|  | Average | Median | Minimum | Maximum | 10th percentile | 90th percentile | Std. deviation |
| --- | --- | --- | --- | --- | --- | --- | --- |
| Commercial to MEC | 364.452 | 368.000 | 201.000 | 384.000 | 350.000 | 377.700 | 16.671 |
| Commercial to cloud | 366.861 | 369.000 | 303.000 | 372.000 | 358.000 | 372.000 | 7.239 |
| Private to MEC | 374.883 | 377.000 | 364.000 | 400.000 | 371.000 | 377.000 | 3.366 |
| Private to cloud | 375.224 | 377.000 | 360.000 | 378.000 | 371.000 | 378.000 | 3.513 |

**TABLE 5** Network uplink throughput measurements (Mbps).

|  | Average | Median | Minimum | Maximum | 10th percentile | 90th percentile | Std. deviation |
| --- | --- | --- | --- | --- | --- | --- | --- |
| Commercial to MEC | 105.209 | 106.000 | 76.800 | 120.000 | 104.000 | 107.000 | 2.796 |
| Commercial to cloud | 100.683 | 101.000 | 31.500 | 110.000 | 99.300 | 102.000 | 3.624 |
| Private to MEC | 56.929 | 56.900 | 54.900 | 59.100 | 56.500 | 57.400 | 0.393 |
| Private to cloud | 56.750 | 56.800 | 5.180 | 58.500 | 56.600 | 57.100 | 1.649 |

Tables 4 and 5 respectively show the downlink and uplink throughput statistics between the UE and the computing platforms, for the different networking and offloading configurations. Average downlink results are very similar in all network setup configurations, which suggests that the capacity is limited by the radio fronthaul. The differences between the maximum data rates of the private network setup and the maximum theoretical values as estimated above can be explained by erroneous transmissions that lead to the use of lower MCS indices. As previously described, the 5G SA private network has 50 MHz of bandwidth whereas the 5G NSA commercial network channel has 60 MHz. Also as previously said, the UE respectively used three and one data layers for the downlink and uplink communications when connected to the private network, and it used two data layers for each channel in the commercial network. The similar downlink data rates of both network setups mean that the higher bandwidth of the commercial network channel compensates for the less data layers, whereas the uplink data rate of the commercial network is slightly less than twice the data rate of the private network, which could be expected given the data layers of the uplink channel. Regarding predictability, the data rates of the private network setup have considerable less standard deviation than the data rates of the commercial network.

## 5.3 | Evaluation of offloading on the 5G network setups

Figure 8 shows the histograms (a) and box plots (b) of the service delay (total delay) and its three components for the commercial 5G network setup when the application is offloaded to a commercial cloud platform. The box plot shows a clear asymmetric distribution of the delay components that depends on network conditions. The median value for the total service delay was 140.117 ms, with a standard deviation of 7.226 ms and 80% of the measurements in [135.029, 152.849] ms. The main contribution to total delay was processing time, which had a median value of 55.696 ms and a standard deviation of 4.362 ms. The image and command transmission times had respective median values of 44.163 and 39.067 ms and standard deviations of 5.266 and 1.385 ms.

Figure 9 shows the same measurements for the commercial network setup when the application is offloaded to its own MEC platform. In this case, the median value of total service delay was noticeably lower, 68.414 ms. Nevertheless, is worth mentioning that a significant number of frames required very long transmission times, up to 323.033 ms, resulting in a worst-case total delay of 394.971 ms. Thus, this is the most unpredictable scenario, with an standard deviation of service delay of 25.420 ms. Nevertheless, 80% of the measurements of the total delay were in the [63.467, 79.904] ms range, that is, 72 ms less than with the commercial cloud.

Figure 10 shows the private 5G network scenario when the service is offloaded to the commercial cloud. The service delay lies between the values obtained in the commercial network setups, but it is much more predictable. Total service delay and image and command transmission times had respective median values of 105.632, 25.594, and 22.259 ms. As it could be expected, the image processing times were similar to those of the commercial 5G network setup with a cloud platform, with median value of 57.091 ms and standard deviation of 5.152 ms. This means that the delay gap between the private and commercial network setups with cloud platforms was due to communication delays. The extra communication delay was evenly distributed between image and command transmission times. In terms of predictability, this private network setup had respective standard deviations of 2.379 and 0.993 ms for the transmission of images and commands, which are much lower than the 5.266 and 1.385 ms of the commercial network setup. As a result, total service delay takes values in a much narrower range than in the commercial network setup with a commercial cloud, with 80% of the values in [101.698, 114.018] ms.

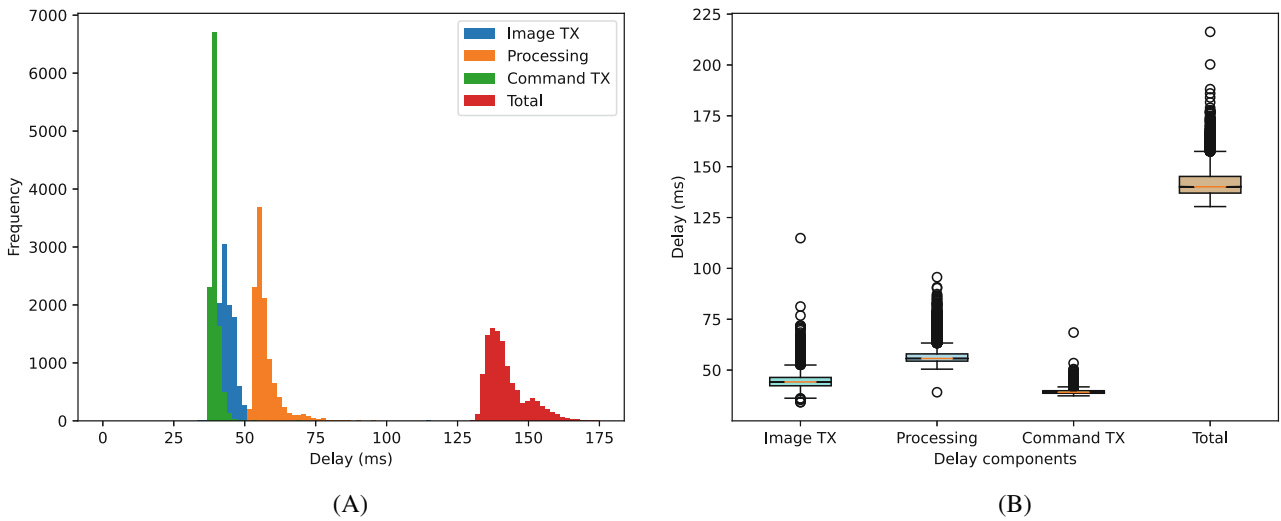

**FIGURE 8** Service delay, commercial 5G network setup with commercial cloud. (A) Histograms; (B) box plots

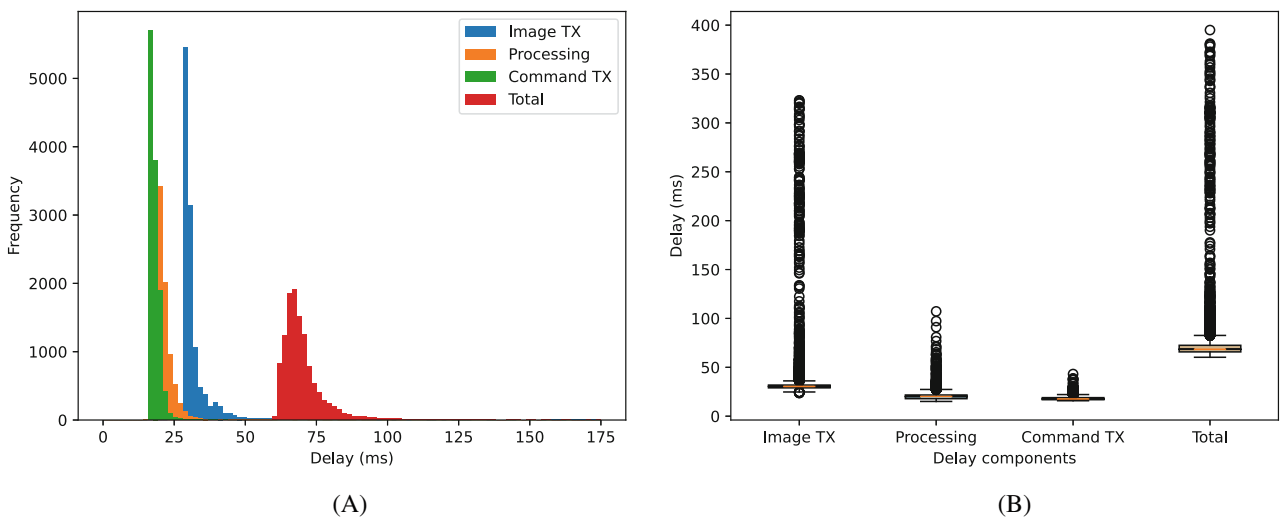

**FIGURE 9** Service delay, commercial 5G network setup with commercial MEC platform. (A) Histograms; (B) box plots

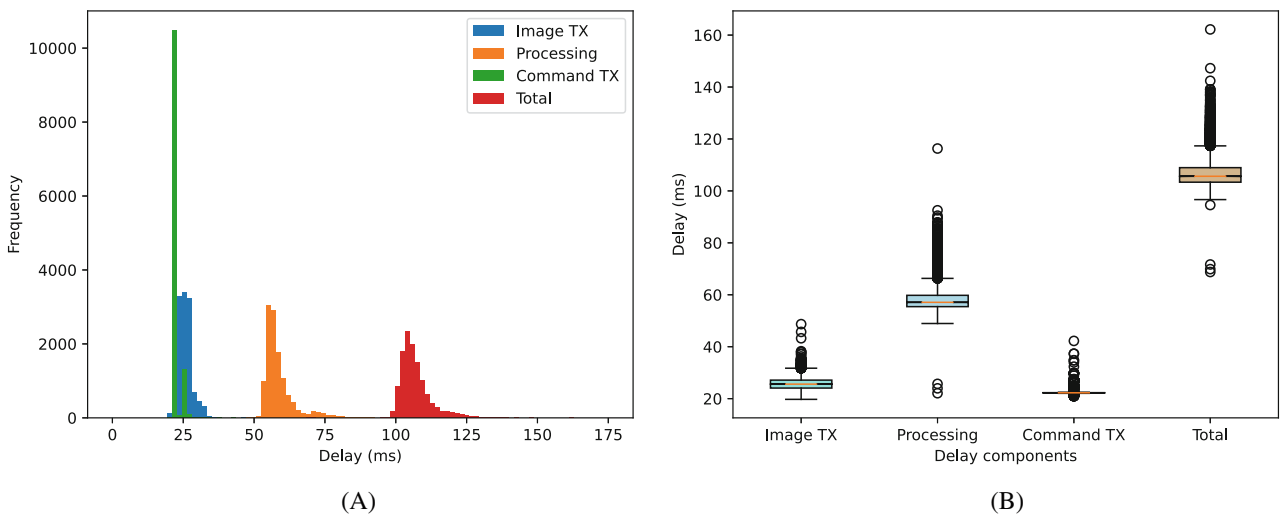

**FIGURE 10** Service delay, private 5G network setup with commercial cloud. (A) Histograms; (B) box plots

Finally, Figure 11 shows the results for the private 5G network with dedicated MEC resources. The median value for total service delay dropped to 57.052 ms in this case, between a 10th percentile of 53.337 ms and a 90th percentile of 61.949 ms. Note that, regardless of the apparently worse specifications of the commercial MEC platform in Table 2, image processing times were longer on the private MEC platform (median value of 34.942 ms compared to only 20.498 ms). This may be due to optimizations of the commercial MEC that are unknown to us, but, regardless of the reason, using the same computing platform in this scenario would result in an extra service delay reduction of almost 15 ms. It can also be observed that, due to the network proximity of all the elements in the setup, transmission delays were much lower in the private network with MEC offloading: the median values of image and command transmission times were 12.149 and 9.867 ms, respectively.

For the sake of clarity, Tables 6–9 summarize the numerical results of the evaluation.

## 5.4 | Evaluation of offloading in WLAN networks

Previous research has emulated proximity computing scenarios using Wi-Fi WLANs. Therefore, for comparison with such a setup, we evaluated the offloading of the target application to a server in a Wi-Fi testbed. Figure 12 shows its architecture.

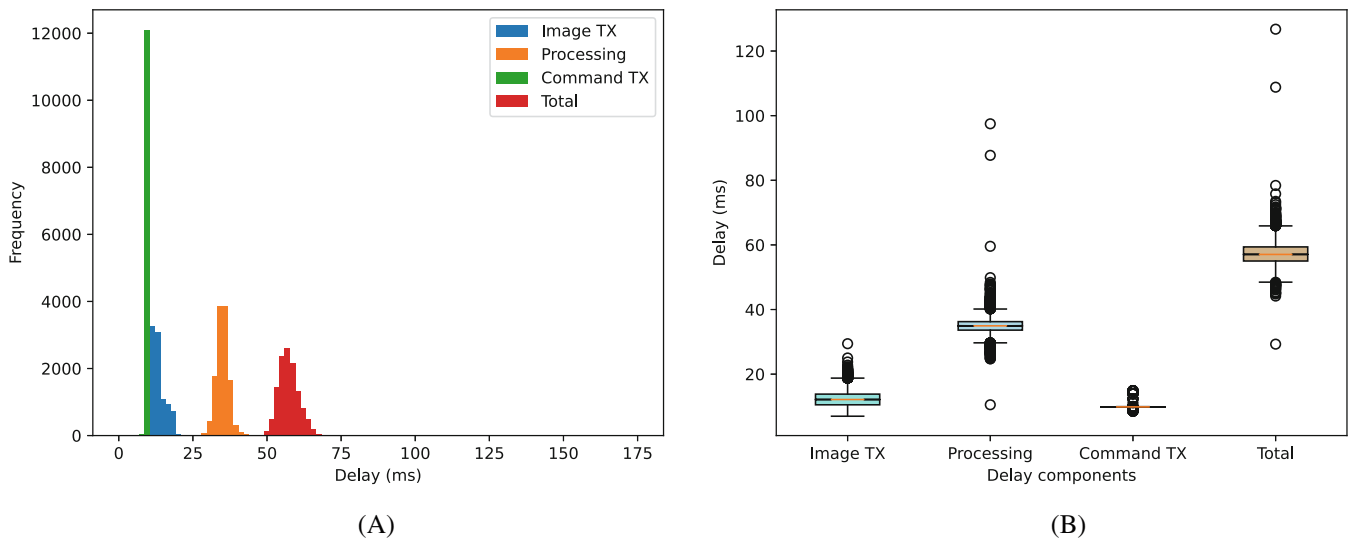

**FIGURE 11** Service delay, private 5G network setup with private MEC resources. (A) Histograms; (B) box plots

**TABLE 6** Image transmission time results (ms).

|  | Average | Median | Minimum | Maximum | 10th percentile | 90th percentile | Std. deviation |
|---|---|---|---|---|---|---|---|
| Commercial to cloud | 45.684 | 44.163 | 34.147 | 114.861 | 41.400 | 56.400 | 5.266 |
| Private to cloud | 25.817 | 25.594 | 19.733 | 48.692 | 23.229 | 28.675 | 2.379 |
| Commercial to MEC | 34.409 | 30.407 | 23.677 | 323.033 | 28.725 | 38.323 | 23.866 |
| Private to MEC | 12.570 | 12.149 | 6.981 | 29.423 | 9.545 | 16.695 | 2.604 |

**TABLE 7** Command transmission time results (ms).

|  | Average | Median | Minimum | Maximum | 10th percentile | 90th percentile | Std. deviation |
|---|---|---|---|---|---|---|---|
| Commercial to cloud | 39.454 | 39.067 | 37.328 | 68.459 | 38.197 | 41.297 | 1.385 |
| Private to cloud | 22.554 | 22.259 | 20.908 | 42.225 | 22.159 | 24.693 | 0.993 |
| Commercial to MEC | 18.062 | 17.617 | 15.783 | 43.103 | 16.405 | 20.121 | 1.721 |
| Private to MEC | 9.841 | 9.867 | 8.555 | 14.919 | 9.786 | 9.916 | 0.321 |

**TABLE 8** Image processing time results (ms).

|  | Average | Median | Minimum | Maximum | 10th percentile | 90th percentile | Std. deviation |
|---|---|---|---|---|---|---|---|
| Commercial to cloud | 56.987 | 55.696 | 39.103 | 95.682 | 53.504 | 61.794 | 4.362 |
| Private to cloud | 58.580 | 57.091 | 22.092 | 116.320 | 54.377 | 64.343 | 5.152 |
| Commercial to MEC | 20.498 | 20.135 | 15.064 | 107.210 | 16.986 | 24.265 | 3.974 |
| Private to MEC | 34.942 | 34.951 | 10.528 | 97.481 | 32.396 | 37.348 | 2.274 |

**TABLE 9** Service delay results (ms).

|  | Average | Median | Minimum | Maximum | 10th percentile | 90th percentile | Std. deviation |
|---|---|---|---|---|---|---|---|
| Commercial to cloud | 142.125 | 140.117 | 130.38 | 216.277 | 135.029 | 152.849 | 7.226 |
| Private to cloud | 106.991 | 105.632 | 68.768 | 162.182 | 101.698 | 114.018 | 5.669 |
| Commercial to MEC | 72.970 | 68.414 | 60.251 | 394.971 | 63.467 | 79.904 | 25.420 |
| Private to MEC | 57.362 | 57.052 | 29.283 | 126.764 | 53.337 | 61.949 | 3.477 |

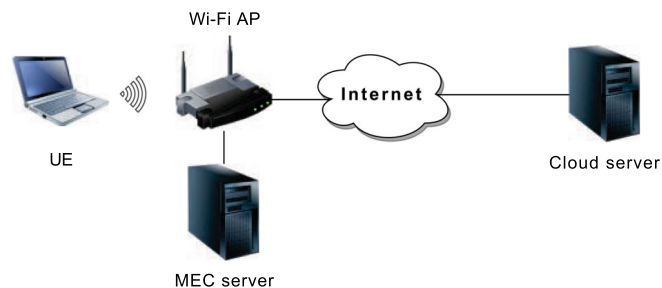

**FIGURE 12** Architecture of the WLAN setup.

The UE was connected wirelessly to an ASUS RT-AX88U[††] Wi-Fi Access Point (AP), which was directly connected to the MEC server through an Ethernet wire. The AP was also connected to the Internet to reach the cloud server. The UE had an Intel Wi-Fi 6 AX201 Wi-Fi module.[‡‡] We followed the same methodology as in the 5G network tests, by measuring the instants when the UE and the MEC server transmitted image and command transmission packets.

Table 10 shows the results. As expected, a Wi-Fi WLAN setup performs much better than a 5G network setup when the UE does not share the channel with any other devices. Delays are lower and much more predictable. Since cellular transmissions are scheduled, they can be delayed to specific times. On the contrary, Wi-Fi transmissions take place as soon as the channel becomes idle after a backoff time, and this is beneficial for an ideal scenario with a single UE. Nevertheless, the Wi-Fi contention method may also lead to much poorer delays when many end devices coexist within the same channel. To achieve higher data rates, Wi-Fi devices usually aggregate multiple layer-3 packets into a long transmission burst, at the expense of longer latency. Moreover, even if the end devices of the network are not configured to use long buffers, they can still experience long delays due to the behaviour of other wireless devices within the same license-free channel. Indeed, Wi-Fi devices may use of the channel for up to 32.767 ms.[40] In addition, cellular networks adapt their modulation and coding rate better to dynamic channel conditions and are not subject to the strict constraints of industrial, scientific, and medical (ISM) bands.

---

[††]https://www.asus.com/es/Networking-IoT-Servers/WiFi-Routers/ASUS-Gaming-Routers/RT-AX88U/
[‡‡]https://www.intel.es/content/www/es/es/products/sku/130293/intel-wifi-6-ax201-gig/specifications.html

**TABLE 10** Transmission delay on a WLAN testbed (ms).

|  | Average | Median | Minimum | Maximum | 10th percentile | 90th percentile | Std. deviation |
| --- | --- | --- | --- | --- | --- | --- | --- |
| Image transmission time, MEC | 1.011 | 0.884 | 0.545 | 3.214 | 0.820 | 1.683 | 0.378 |
| Command transmission time, MEC | 0.898 | 0.515 | 0.472 | 4.373 | 0.479 | 2.053 | 0.624 |
| Image transmission time, cloud | 13.477 | 13.343 | 11.687 | 21.342 | 13.260 | 14.162 | 0.398 |
| Command transmission time, cloud | 13.247 | 12.886 | 12.759 | 16.725 | 12.810 | 14.407 | 0.625 |

## 5.5 | Onboard versus offload computing

To assess the interest of MEC offloading versus onboard computing, we evaluated the performance of the application on embedded devices that are suitable for AI operations.

We evaluated different embedded boards that can run the application in our use case and can be installed in a drone (given their power consumption, weight, size, and price). Nowadays two solutions stand out:[41] A low-power computing board, such as a Raspberry Pi, with a neural computing peripheral such as the Intel NCS 2;[§§] and specialised low-consumption boards integrating a GPU for machine learning acceleration, such as the Nvidia Jetson boards.[¶¶] In this regard, the Jetson Nano and Jetson Xavier NX boards are suitable, as they have been used in similar experiments. Table 11 shows their main technical specifications.

The study in Reference 41 compared the performances of an Intel NCS 2, a Jetson Nano and a Jetson Xavier NX for common AI applications. The Jetson boards can execute most general AI models. The Intel NCS 2 only supports OpenVINO models in principle, but the application in this work relies on two OpenVINO models that are well supported by the general purpose computing servers in MEC and cloud platforms.

Table 12 shows the processing time of the target application, that is, the time it takes to generate a result from the moment the image is taken by the camera, in the most stringent layout with an Intel NCS 2 connected to a Raspberry Pi 4. Average processing time was 216.398 ms, for an average frame rate of 4.621 frames per second. It is worth mentioning that 96.55% of the samples took less than 243.126 ms, but 3.45% of the samples took much longer, between 1021.061 and 1062.580 ms, due to few instances of the re-identification task. We verified that this was not due to high CPU usage nor CPU throttling, and memory usage was also well under maximum capacity. According to Reference 41, for the YOLOv3 model, which is equivalent in complexity to our application, the Jetson Nano has 0.64–0.68 times the performance of a Raspberry Pi 4 with an Intel NCS 2, whereas the Jetson Xavier NX outperforms the Intel NCS 2 by a 2.36–2.44 factor. We therefore estimate that the Jetson Nano and Jetson Xavier NX could respectively process 3 and 11 frames per second on average with AI models like ours in similar conditions.

**TABLE 11** UAV onboard computing solutions for AI applications.

|  | Nominal AI performance (TFLOPS) | Weight (grams) | Size (mm) | Nominal power consumption (W) | Price ($) |
| --- | --- | --- | --- | --- | --- |
| Raspberry Pi 4 + Intel NCS 2 | 1[41] | 46 + 18 | 153 × 56 × 19 | 6.25 + 1 | 45 + 99 |
| Nvidia Jetson Nano | 0.472 | 157 | 100 × 80 × 29 | 10 | 99 |
| Nvidia Jetson Xavier NX | 1.33 | 172 | 103 × 90.5 × 34 | 15 | 399 |

§§https://www.intel.es/content/www/es/es/products/sku/140109/intel-neural-compute-stick-2/specifications.html
¶¶https://developer.nvidia.com/embedded/jetson-modules

**TABLE 12** Processing time in Intel NCS 2 (ms).

| Average | Median | Minimum | Maximum | 10th percentile | 90th percentile | Std. deviation |
| --- | --- | --- | --- | --- | --- | --- |
| 216.398 | 177.420 | 172.979 | 1062.580 | 175.739 | 226.931 | 157.276 |

In terms of energy consumption, on the one hand, the average root mean square (RMS) power consumption of an onboard computing solution consisting of a Raspberry Pi 4, an Intel NCS 2 and a Logitech StreamCam web camera is around 5.68 W with our application. During offloading evaluations, the power consumption of the payload was only around 4.05 W. Of these, 0.97 W corresponded to the Logitech Streamcam web camera, which was also used for onboard computing. In Reference 41 the authors reported respective average power consumptions of 7.95 and 15.2 W for the Nvidia Jetson Nano and Xavier NX boards running a YOLOv3 object detection model. On the other hand, the DJI Matrice M210 drone in our tests has a reported flight time of 23 minutes without any payload using two regular DJI TB50 batteries. These batteries have a nominal capacity of 4280 mAh for 22.8 V, which results in a stored energy of 195.17 Wh. By considering the energy of the batteries and the reported flight time, the average power consumption of the drone without any payload is 509.13 W.

Table 13 shows the energy consumption of the payload components and the estimated flight time versus the case without any drone payload. This is an upper bound, as the motors of the drone spend additional power due to the extra weight of the payload, but the range extension due to offloading seems to be small. Therefore, an discussed in Section 6.6, the main justification for computing offloading is the increased processing power that makes more application scenarios feasible, although the trade-off between the costs of extra onboard hardware and remote computing services should also be taken into account.

## 5.6 | Discussion

The evaluations of 5G offloading network setups, summarised in Tables 6–9, reveal some interesting facts.

First, it is essential to distinguish between communication and processing delays. The latter only depend on the resources available and the current demand of the offloading platforms, regardless of the location of the processing platform at the cloud or the edge of the network. As an example, even though the contributions to total service delay of the image and transmission times in Tables 6 and 7 are dominant in our case, if the cloud platform and the MEC platform of the private network had similar processing capabilities as those of the MEC platform of the commercial network, total service delay would be 35 ms lower on average in the cloud scenarios and almost 15 ms lower in the private network setup with MEC, yielding reductions above 25% in total service delay in the commercial network setup with cloud, the private network setup with cloud, and the private network setup with MEC. This means, therefore, that the effect of the computational capacity of the offloading platform on application performance should not be neglected. Logically this effect will depend on the computational cost of the application. The greater the cost, the greater the effect.

Regarding transmission delays, the results show considerable differences in total service delay in a given network setup when using MEC or cloud, highlighting the importance of the geographical proximity of the offloading server to the end user to achieve high performance. The differences in image transmission delays are more noticeable between commercial and private network setups than between cloud and MEC scenarios. Even though in our evaluations other external UEs coexisted in the commercial network with our end user, the differences suggest a longer data path between the terminal in the tests and the offloading platform. That is, whereas in the private network the 5G gNodeB is directly connected to the UPF, which is in turn directly connected to the Internet, it is very likely that in the commercial network the UPF is much farther away from the gNodeB, and many forwarding nodes exist between these elements. In contrast to processing delays, the transmission delays of cellular networks are less dependent on the target application, owing to their scheduling mechanisms, assuming that they can provide enough data rate to the UE. This is not the case in other access networks, whose transmission delays depend primarily on packet payload size, network data rate, and channel occupancy time.

**TABLE 13** Comparison of energy consumption of onboard computing and offloading solutions.

| | Payload elements | Weight (grams) | Power consumption of payload (W) | Estimated flight time versus no payload |
| --- | --- | --- | --- | --- |
| Offloading | Pi 4 + Streamcam | 196 | 4.05 | 99.20% |
| Onboard, NCS 2 | Pi 4 + NCS 2 + Streamcam | 214 | 5.68 | 98.91% |
| Onboard, Nano | Nano + Streamcam | 307 | 7.95[41] | 98.43% |
| Onboard, Xavier NX | Xavier NX + Streamcam | 322 | 15.2[41] | 97.01% |

In addition to average communication delays, a key parameter to evaluate the performance of the offloading scenarios is delay predictability, by checking the proportion of transmissions whose delays differ significantly from typical values. This deviation will be more or less relevant depending on the use case. While certain critical applications may not be able to withstand any long delays at all, other applications may be tolerant to such delays or even to transmission losses. In this regard, the private network was considerably more predictable, specially with its MEC offloading platform. Note the unacceptable maximum image transmission delay of 323.033 ms in the commercial network setup with a MEC platform. Although 90% of the image transmission delays in this case were below 126% of the median value and 97.935% were below twice the median value, over 1% of the image transmissions took longer than 189.475 ms. We believe this is due to the immaturity of the commercial network with MEC, so these outliers may be less severe in the future, and the 90th percentile results may be more representative.

Previous research has relied on Wi-Fi WLANs to emulate proximity computing scenarios. As a reference, in Section 6.4 we have evaluated image and command transmission times when the application in our use case is offloaded to MEC or cloud servers through a Wi-Fi access network. The evaluation was conducted under the same premise of no other external devices sharing the same channel. In this setup, the Wi-Fi interface hardly introduces any delay. Image and command transmission times using MEC or the cloud cannot be extrapolated to a real 5G MEC scenario. According to the results in Table 10, a Wi-Fi-based offloading setup would be feasible in all the scenarios of our use case, even if the application is offloaded to a cloud server. Nevertheless, these results would not be achievable in a more open general setup, in which multiple external heterogeneous devices would coexist within the same license-free channel. Wi-Fi transmissions employing packet aggregation techniques to share the channel more efficiently can take up to 32.767 ms.[40] Therefore, delays may be much less predictable in reality than in tightly controlled laboratory environments. Moreover, cellular networks have additional capabilities over a Wi-Fi access network beyond transmission scheduling, such as improved adaptive modulations.

Regarding the feasibility of the target application, if we compare the 90th percentile service delays in Table 9 with the service delay threshold references $t_w$, $t_r$, and $t_b$ defined in Section 6.1, the application can produce motion corrections before a walking person covers one third of the distance to the edge of the frame in all the offloading network setups. A running person could be tracked in all these setups in most cases, but maximum total service delay would be excessive in the commercial network setups. The same would happen with cyclists in the commercial 5G network with MEC: 6.42% of the total service delay measurements in the experiments exceeded the $t_b$ threshold. This probability was only 0.02% in the private 5G network setup with MEC.

In Section 6.5, as an alternative to computing offload, we have evaluated commercial embedded boards with capabilities to run AI applications onboard. For the particular use case in this work, energy consumption has practically no impact on UAV flight time. All embedded boards considered would fulfil the requirements to track a walking person. A runner would not be reliably tracked with the Intel NCS 2, as 21.26% of the processing times exceeded the target threshold $t_r$. The Nvidia Jetson Nano is even more limited. Compared to the 5G network offloading scenarios, the performance of the Nvidia Jetson Xavier NX would lie between those of the MEC and cloud setups. According to our estimations, it would track a runner satisfactorily, but, given its processing times, it would not be able to track a cyclist.

Table 14 shows feasible and non feasible scenarios using application offloading and onboard computing, based on the 90th percentile results of the evaluations. All 5G network setups with MEC are feasible according to this criterion.

Summing up, real 5G networks with MEC resources can support scenarios of our use case that would be unfeasible with cloud computing. Moreover, offloading has higher, more predictable performance in 5G private networks than in commercial networks. It is also interesting to note that, even though the minimum delays in Tables 6 and 7 are consistent with the expected best-case delays of an ideal 5G network (below 10 ms), in a real private network with dedicated MEC resources *typical* delays are considerably higher. Finally, even though the power consumption of AI-enabled commercial embedded boards has low impact on UAV flight time, they cannot fulfil the requirements of computationally demanding applications. Thus, UAV computing offloading is currently useful for practical use cases like ours.

**TABLE 14** Feasibility of application offloading and onboard computing scenarios (considering 90th percentile total service times).

|  | Walking | Running | Cycling |
| --- | --- | --- | --- |
| Threshold | 419.2 | 209.6 | 83.84 |
| 5G, commercial to cloud | 152.849 | 152.849 | 152.849 (NF) |
| 5G, private to cloud | 114.018 | 114.018 | 114.018 (NF) |
| 5G, commercial to MEC | 79.904 | 79.904 | 79.904 |
| 5G, private to MEC | 61.949 | 61.949 | 61.949 |
| Onboard Intel NCS 2 | 226.931 | 226.931 (NF) | 226.931 (NF) |

*Note*: Times in ms.
Abbreviation: NF: not feasible.

## 6 | CONCLUSIONS

Aiming at supporting new use cases that require wireless connectivity under very stringent quality of service requirements, 5G networks have embraced new technologies and paradigms such as edge computing. By bringing computational resources closer to the end users, edge computing can improve the time responses, data rates, predictability, and privacy of the network services. This is specially interesting for many UAV applications that are computationally intensive and require low response times.

In this article, we have evaluated the first experience of application offloading from drones to real 5G networks with commercial and private carrier-grade MEC systems. A follow-me drone service has been implemented as a representative use case. We have characterised this application and evaluated its latency in a NSA 5G operator network and a SA 5G carrier-grade private network, by considering both MEC and cloud offloading scenarios. The results suggest that a cloud-assisted scenario cannot be completely ruled out, but a MEC-assisted scenario is necessary to provide good quality results and guarantee the feasibility of the application when the user is moving fast (e.g., riding a bicycle). Moreover, we conducted the same evaluations in WLAN-based offloading setups and on AI-enabled embedded boards that can be installed in an UAV. the results show that Wi-Fi offloading setups provide an upper performance bound in optimal environments, whereas currently available embedded boards cannot fulfil the requirements of the most demanding scenarios of our use case. We conclude that dedicated MEC resources at the gNodeBs would be necessary for some applications with stringent latency requirements, since otherwise the variability of service delays is too high, and maximum transmission delays exceed expected 5G references by large.

**AUTHOR CONTRIBUTIONS**
**David Candal-Ventureira:** conceptual & experimental design and implementations. **Francisco Javier González-Castaño:** conceptual & experimental design. **Felipe Gil-Castiñeira:** conceptual design and hardware. **Pablo Fondo-Ferreiro:** experimental design and 5G architecture.

**ACKNOWLEDGMENTS**
This research has been partially supported by Spanish Grants PDC2021-121335-C21, PID2020-116329GB-C21, and PRE2021-098290, funded by MCIN/AEI/10.13039/501100011033 and FSE+. Open access funded by University of Vigo.

**DATA AVAILABILITY STATEMENT**
The data that support the findings will be available upon request to the authors due to industry contract.

**ORCID**
*Felipe Gil-Castiñeira*  https://orcid.org/0000-0002-5164-0855